\documentclass{ws-ijmpa}

\usepackage[super]{cite}
\usepackage{xcolor}
\usepackage{hyperref}

\hypersetup{colorlinks=false,allbordercolors=blue,pdfborderstyle={/S/U/W 1}}
\usepackage{amsmath}
\usepackage{mathtools}
\usepackage{amssymb}
\usepackage{slashed}
\usepackage{braket}
\usepackage{bm}
\usepackage{float}
\usepackage{revsymb}
\usepackage{physics} 
\usepackage{siunitx} 
\usepackage{enumerate} 

\usepackage{graphicx}
\usepackage{pgfplots}
\usepackage{pgfplotstable}
\usepackage{tikz,pgfplots}

\usepackage{booktabs}
\usepackage{subcaption}

\pgfplotsset{compat=1.14}
\begin{document}

\markboth{A. F. Candar, M. C. Güçlü}{Electron positron pair production accompanied by GDR}

\catchline{}{}{}{}{}
%
%%%%%%%%%%%%%%%%%%%%%%%%%%%%%%%%%%%%%%%%%%%%%%%%%%%%%%%%%%%%%%%%%%%%

\title{Electron positron pair production accompanied by giant
dipole resonance at RHIC and LHC}

\author{A. F. Candar}

\address{Physics Department, Istanbul Technical University, \\
Maslak 34469 Istanbul, Türkiye\\
candara15@itu.edu.tr}

\author{M. C. Güçlü}

\address{Physics Department, Istanbul Technical University, \\
Maslak 34469 Istanbul, Türkiye\\
guclu@itu.edu.tr}

\maketitle

\begin{history}
\received{Day Month Year}
\revised{Day Month Year}
\accepted{Day Month Year}
\published{Day Month Year}
\end{history}

\begin{abstract}
In ultra-peripheral heavy ion collisions (UPCs), the broad energy spectrum of photons enables a variety of physical processes, ranging from lepton pair production to giant dipole resonances (GDRs). In this work, we evaluate production of electron-positron pair production cross section with GDR. Specifically, we focus on Au+Au collisions at a center of mass energy of $\sqrt{s_{NN}}=$ 200 GeV and Pb+Pb collisions of $\sqrt{s_{NN}}=$ 2.76 TeV per nucleon. We calculate cross sections while taking into account with and without the kinematic restrictions relevant to STAR detectors. We also present the differential cross sections in terms of the variables rapidity $(y)$, transverse momentum ($p_\perp$) and invariant mass ($M$) to compare them with respect to RHIC and LHC energies.
\end{abstract}


\ccode{PACS numbers: 25.75.Cj, 25.75.Dw, 25.75.-q}

\section{Introduction}

\begin{figure}[t]
\centering
\includegraphics[width=0.5\textwidth]{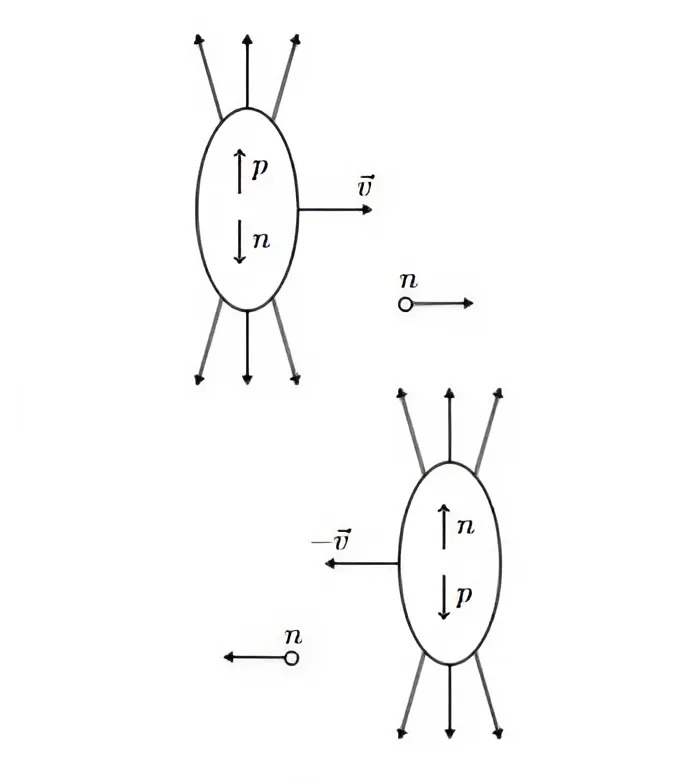}
        \caption{Schematic figure of giant dipole motion in colliding of two relativistic heavy ions. The electric fields of the heavy ions pushes the protons upwards and the neutrons then move downwards. During this motion some neutrons leave the nuclei and move in the beam direction.}
\label{fig:giantdipole}
\end{figure}
    \hfill

Photon-induced reactions give us many opportunities to study a diverse range of physics. Especially, at RHIC and at LHC, ultra-relativistic peripheral collisions of heavy ions can produce copies of numbers of lepton pairs via the two-photon process. The energies of the colliding heavy ions in these colliders start from 100 GeV up to 5000 GeV per nucleon. \cite{adam2018low,dyndal2017electromagnetic,abbas2013charmonium,adam2021measurement} The cross sections of electron-positron pairs are quite large so that it is possible to test QED for the strong field mainly for the ultra-peripheral collisions (UPC). UPCs of heavy ions are described by an impact parameter $b$ greater than the sum of the radius of the colliding nuclei. In this case, nuclei pass each other at impact parameters larger than the sum of their nuclear radii, so they avoid direct overlap between the nuclei. Therefore, the hadronic interactions are eliminated, and electromagnetic interactions become dominant. The intense electromagnetic fields are generated by the high charge numbers of the nuclei (e.g., $Z_{Au}=79$, $Z_{Pb}=82$) at the close to speed of light so that electromagnetic interactions and photon-induced reactions become the dominant focus \cite{bertulani2005physics, baltz2008physics}. 

On the other hand, for central collisions, main study is around the physics of the quark-gluon plasma, a state of matter characterized by deconfined quarks and gluons. 

Photon-photon ($\gamma\gamma$) and photon-nucleus ($\gamma Z$) interactions are occurs commonly in UPC, and studying these interactions provide valuable insights about QED. Photon-photon interactions mainly result in final states such as lepton pairs and meson productions \cite{klein2020photonuclear}. Quasi-real photons produced by the relativistic ions causes the photo-nuclear interactions. GDR which is a collective oscillation of protons against neutrons within the nucleus is the the most common outcome of such interactions.\cite{bertulani1988electromagnetic,bertulani1999microscopic} This interaction is depicted in Fig. \ref{fig:giantdipole} in colliding of two relativistic heavy ions.  The electric fields of the heavy ions pushes the protons upwards and the neutrons then move downwards which often results in the emission of a single or multiple neutrons in the beam directions and these neutrons are detected with high precision using Zero Degree Calorimeters (ZDCs).\cite{pshenichnov1999particle,adams2004production}  GDR peaks at relatively low photon energies ($E_\gamma\approx14$ MeV for gold and lead nuclei).\cite{pshenichnov2001mutual,pshenichnov2011electromagnetic} Neutron emissions from nuclear dissociation could be one neutron $1n1n$ or multiple neutron $XnXn$ from each nucleus.\cite{baltz2009two,harland2023exciting} When the excitations energies of the nuclei is higher, both nuclei emit multiple neutrons. (See also Refs. \citen{baur2003electromagnetic,vidovic1993electromagnetic,klusek2014electromagnetic,braun2014hadronic} for more details about nuclear dissociation.)
 
In this study, we investigate the electron positron pair production process using the lowest order Feynman diagrams. As a result, we have obtained impact parameter dependence cross section expression. By improving the calculation techniques, more detailed and accurate cross-section expressions have been obtained. Determination of impact parameter dependence cross section play important role for the production of an electron-positron pair with mutual Coulomb excitation.

\section{Formalism}
\begin{figure}[t]
\centering
\includegraphics[width=0.7\textwidth]{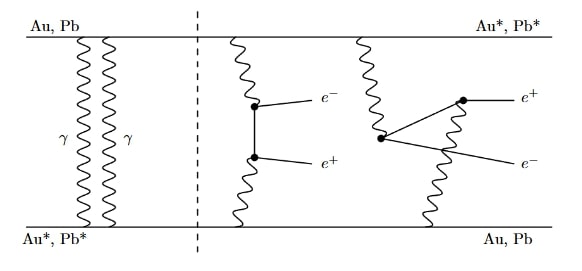}
        \caption{Electron-positron pair production with giant dipole resonance. In this figure only lowest order Feynman diagrams (direct and crossed terms) for electron-positron pair production are shown.}
\label{fig:feynmandiagrams}
\end{figure}

The cross section to produce an electron-positron pair with mutual Coulomb excitation can be written as 

\begin{equation}
	\sigma=\int d^2b \,P_{e^+e^-}(b)\,[P_{GDR}(b)]^2,
\end{equation}

where $P_{e^+e^-}(b)$ is the free pair production probability, $P_{GDR}(b)$ is the the probability of a nucleus being excited by a giant dipole resonance as shown in Fig. \ref{fig:feynmandiagrams}. The task is to determine the each probability accurately. 

    \hfill

We can write the cross section of producing electron-positron pair by integrating the $S$ matrix elements over the impact parameter $b$

\begin{equation}
    \sigma_{e^+e^-} = \int d^2b \sum_{k>0} \sum_{q<0} \left| \langle \chi_k^{(+)} | S_{AB} + S_{BA} | \chi_q^{(-)} \rangle \right|^2,
\end{equation}
where the summation over the states k is restricted to those above the Dirac sea, and the summation over the states q is restricted to those occupied in the Dirac sea. Here $S_{AB}$ is the direct term and  $S_{BA}$ is the exchange term of the $S$ matrix elements in Fig. \ref{fig:feynmandiagrams}. The result for $S_{AB}$ is

\begin{align}
    \langle \chi_k^{(+)} | S_{AB} | \chi_q^{(-)} \rangle =& \frac{i}{2\beta} \int \frac{d^2 \vec{p}_\perp}{(2\pi)^2} \exp \left[ i \left( \vec{p}_\perp - \frac{\vec{k}_\perp + \vec{q}_\perp}{2} \right) \vec{b} \right] \nonumber\\ 
&* \mathcal{F}(\vec{k}_\perp - \vec{p}_\perp; \omega_A) \mathcal{F}(\vec{p}_\perp - \vec{q}_\perp; \omega_B) \mathcal{T}_{kq}(\vec{p}_\perp; \beta),
\end{align}

where the functions $\mathcal{F}(q,\omega)$ is the scalar part of the electromagnetic field  of the moving heavy ions in momentum space and $\mathcal{T}_{kq}(p_\perp; \beta)$ is the propagator of the intermediate lepton and the matrix elements for the coupling of the photon to the leptons.
Here, $\omega_A$ and $\omega_B$ are the frequencies of the virtual photons, $\vec{k}_\perp$ and $\vec{p}_\perp$ are momentums of the electrons and positrons. Also, $\ket{X_k^{(+)}}$ refers to the positive-energy spinors, and $\ket{X_q^{(-)}}$ refers to the negative-energy spinors.

Including both the direct and crossed Feynman diagrams,the results for the cross section,
as a function of impact parameter,can be obtained as
\begin{equation}
    \frac{d\sigma}{db} = \int_0^\infty q dq \, b J_0(qb) F(q).
    \label{eq:dsigmadb}
\end{equation}

Here, $ F(q)$ is a nine-dimensional integral that can be calculated with the Monte Carlo integration method:

\begin{align}
   F(q) =& \frac{\pi}{8\beta^2} \sum_{\sigma_k} \sum_{\sigma_q} \int_0^{2\pi} d\phi_q \int \frac{ dk_z dq_z d^2 k_\perp d^2 K d^2 Q} {(2\pi)^{10}} \Bigl\{ \mathcal{F}(\frac{\vec{Q}-\vec{q}}{2}; \omega_A) \mathcal{F}(-\vec{K}; \omega_B) \nonumber \\ 
   &*\mathcal{T}_{kq}(\vec{k}_\perp - \frac{\vec{Q}-\vec{q}}{2}; \beta) 
    + \mathcal{F}(\frac{\vec{Q}-\vec{q}}{2}; \omega_A) \mathcal{F}(-\vec{K}; \omega_B) \mathcal{T}_{kq}(\vec{k}_\perp - \vec{K}; -\beta)\Bigl\} \nonumber\\
    &*\Bigl\{\mathcal{F}(\frac{\vec{Q}+\vec{q}}{2}; \omega_A) \mathcal{F}(-\vec{q}-\vec{K}; \omega_B) \mathcal{T}_{kq}(\vec{k}_\perp - \frac{\vec{Q}+\vec{q}}{2}; \beta) \nonumber \\
    &+ \mathcal{F}(\frac{\vec{Q}+\vec{q}}{2}; \omega_A) \mathcal{F}(-\vec{q}-\vec{K}; \omega_B) \mathcal{T}_{kq}(\vec{k}_\perp +\vec{q}-\vec{K}; -\beta)\Bigl\}. 
\end{align} 

\begin{figure}[h]
    \centering
    \begin{subfigure}[b]{0.45\textwidth}
        \centering
        \includegraphics[width=\textwidth]{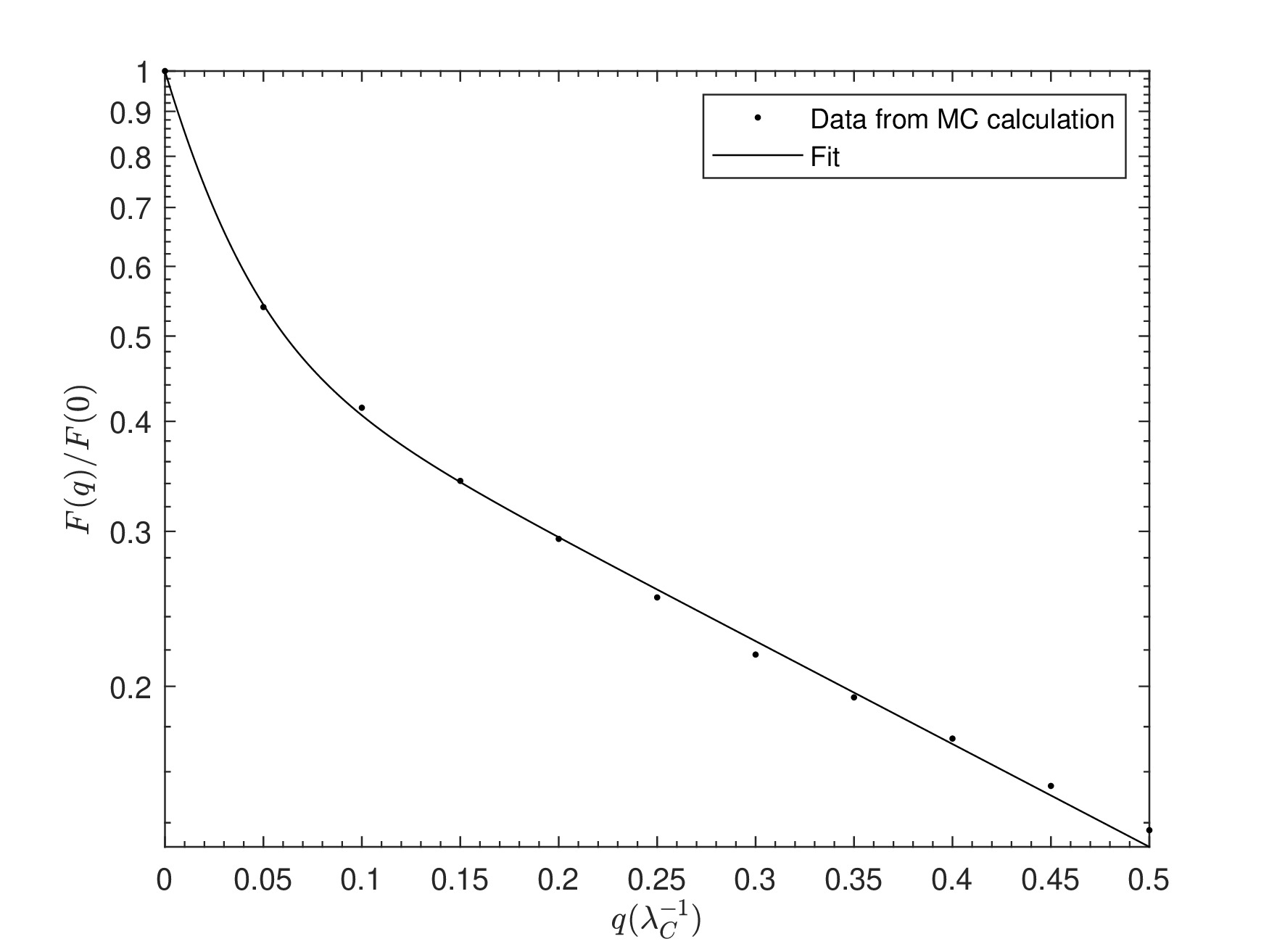}
        \caption{RHIC 200 GeV without restriction}
        \label{fig:grafik1}
    \end{subfigure}
    \hfill
    \begin{subfigure}[b]{0.45\textwidth}
        \centering
        \includegraphics[width=\textwidth]{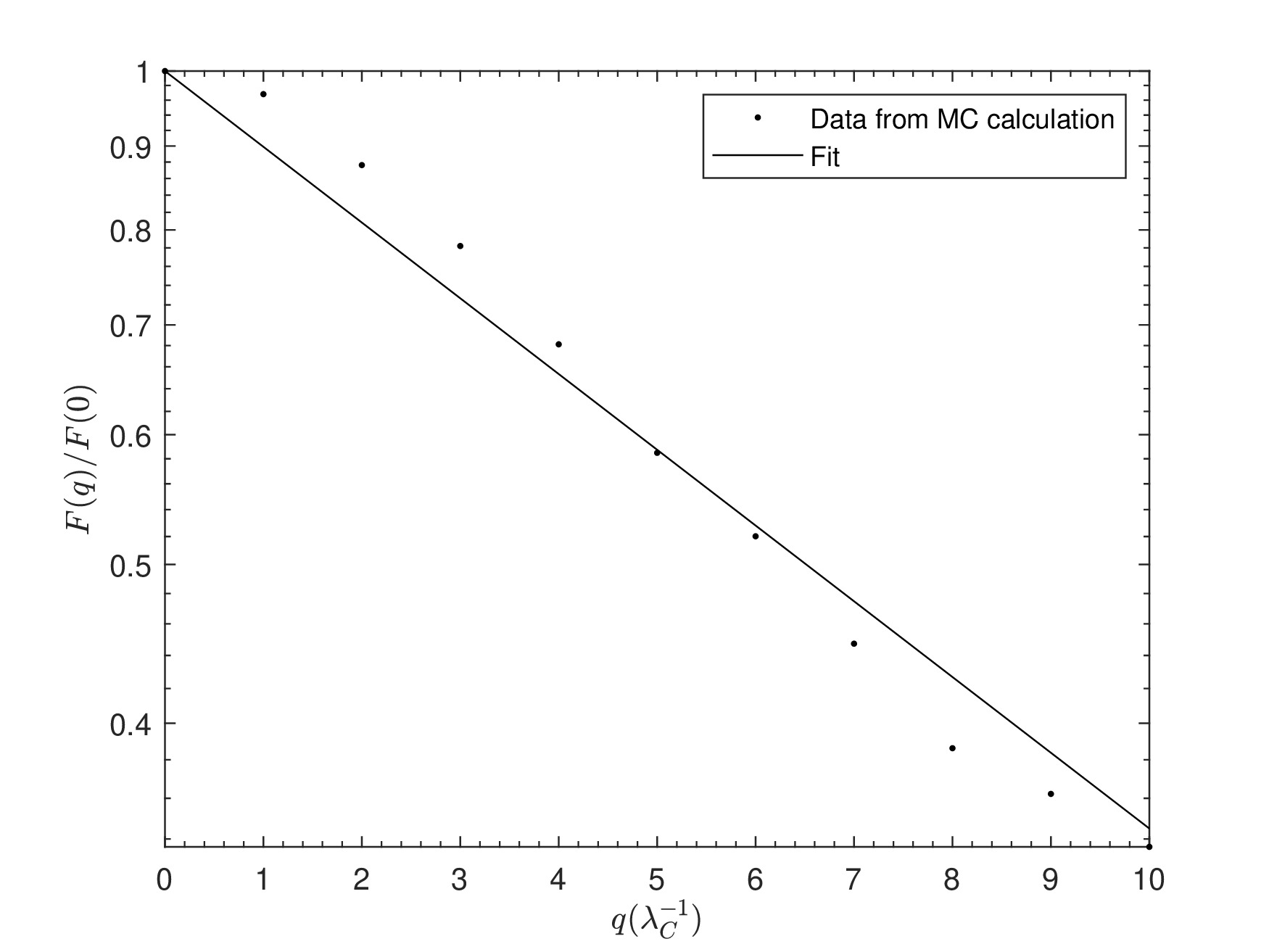}
        \caption{RHIC 200 GeV with restriction}
        \label{fig:grafik2}
    \end{subfigure}
    \vskip\baselineskip
    \begin{subfigure}[b]{0.45\textwidth}
        \centering
        \includegraphics[width=\textwidth]{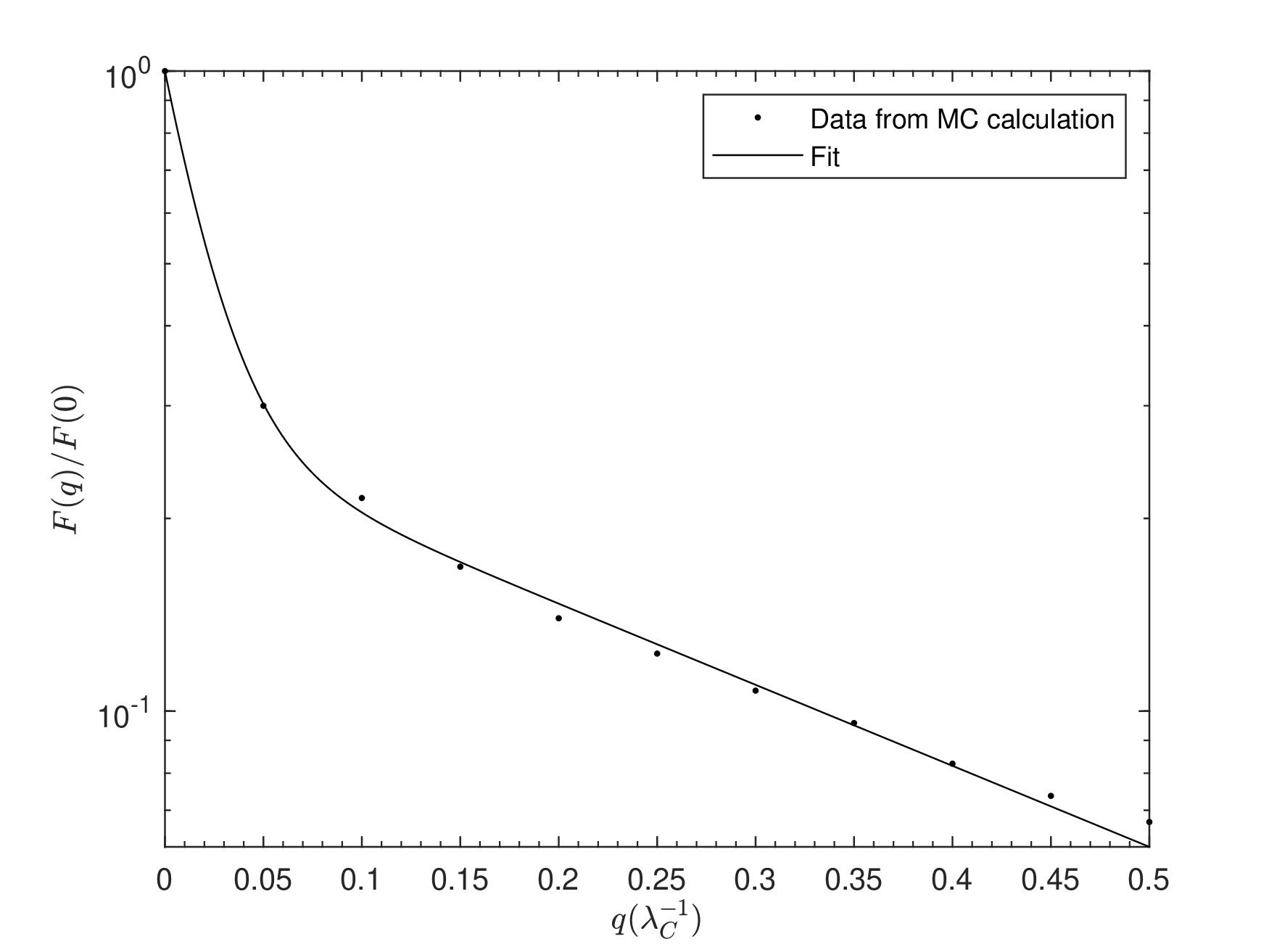}
        \caption{LHC 2.76 TeV without restriction}
        \label{fig:grafik3}
    \end{subfigure}
    \hfill
    \begin{subfigure}[b]{0.45\textwidth}
        \centering
        \includegraphics[width=\textwidth]{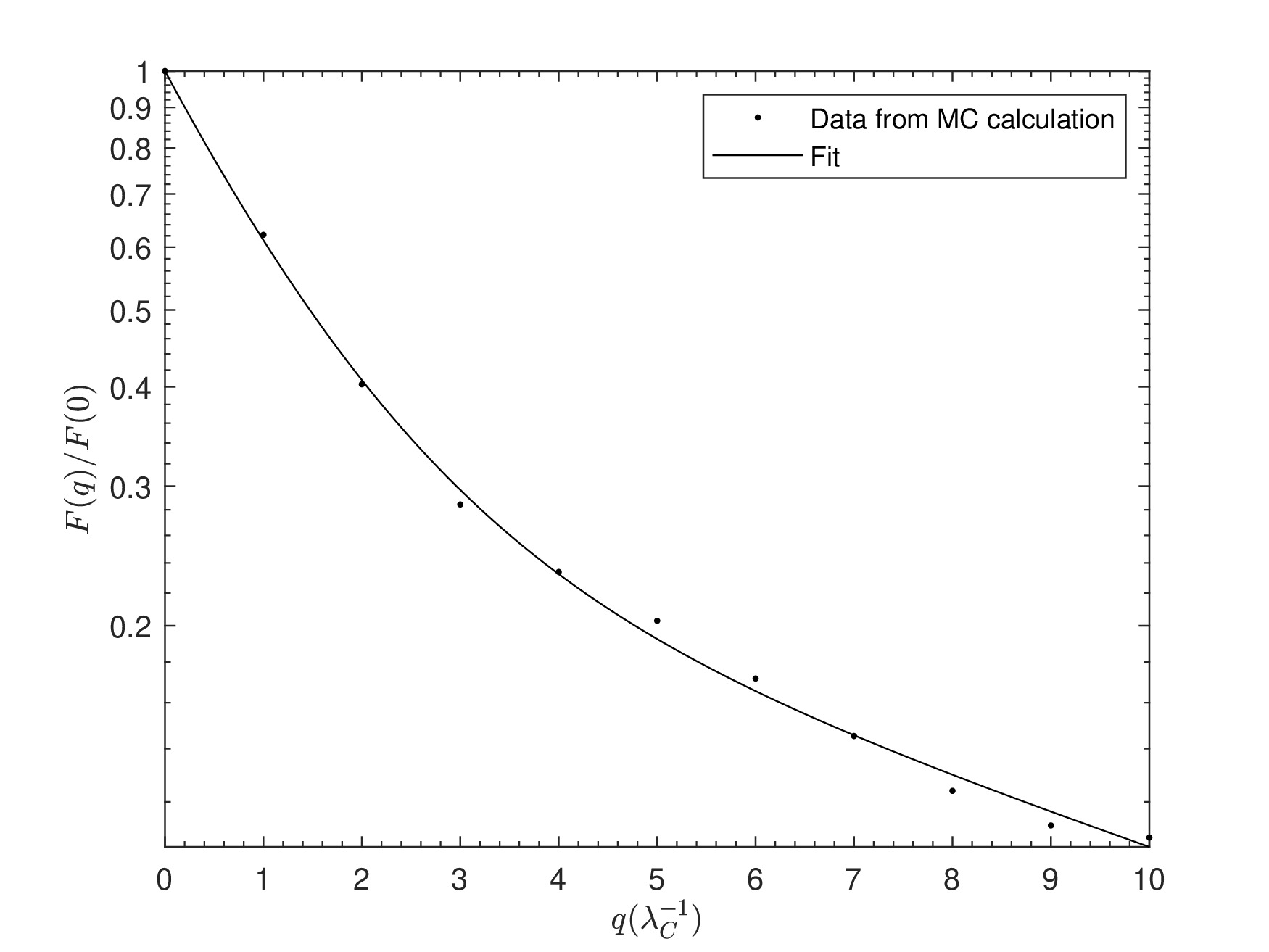}
        \caption{LHC 2.76 TeV with restriction}
        \label{fig:grafik4}
    \end{subfigure}
    \caption{The function $F(q)/F(0)$ is calculated for the RHIC and LHC energies with and without restrictions. The points show the results of the Monte Carlo
calculations for each q value and the smooth curve is our fit for these
points.}
    \label{fig:MCfit}
\end{figure}

We have separately calculated $F(q)$ function since it can not be integrated together with the highly oscillating Bessel function especially for large values of the impact parameters. We have employed Monte Carlo method to calculate the $F(q)$ function and we have used sufficiently large number of Monte Carlo points to achieve a good accuracy. We have plotted these functions for different heavy ions in Fig. \ref{fig:MCfit}. We have done all these calculations for RHIC and LHC energies with and without kinematic restrictions. In our previous works,\cite{gucclu2008lepton,csengul2011bound} these smooth fits were generally in the form of $e^{-aq}$. However, when we improved our calculations, we have found that the most appropriate fits should be in the form of
\begin{equation}
	F(q)=F(0)[(1-a)e^{-bq}+ae^{-cq}]
    \label{eq:F(q)},
\end{equation}

where $F(0)$ gives the total cross section $\sigma_T$ at q = 0. The parameters $a$,$b$ and $c$ are obtained from the suitable fit functions.

To calculate the differential cross section, we can insert the fit function obtained in Eq. \ref{eq:F(q)} into Eq. \ref{eq:dsigmadb}
\begin{equation}
    \frac{d\sigma}{db} = F(0)\int_0^\infty dq  \, qb J_0(qb) [(1-a)e^{-bq}+ae^{-cq}],
\end{equation}

and by using this equation the we can calculate probability of producing electron-positron pairs and it is equal to 

\begin{equation}
    P_{e^+e^-}(b)=\frac{1}{2\pi b }\frac{d\sigma}{db}.
\end{equation}

On the other hand, the probability of Coulomb excitation leading to the emission of a single neutron is given by:
\begin{equation}
    P_{(1n)}(b) = P_{(1n)}^1(b) e^{-P^1(b)},
\end{equation}

here, $P_{(1n)}^1(b)$ represents the initial Coulomb excitation probability at a given impact parameter $b$. This probability is defined for the case where only a single excitation occurs, without considering higher-energy excitations. The initial excitation probability is expressed as:
\begin{equation}
    P^1(b) = \frac{S}{b^2},
\end{equation}

with
\begin{equation}
    S = \frac{2 \alpha^2 Z^3 N}{A m_N \omega} \approx 5.45 \times 10^{-5} Z^3 N A^{-2/3} \text{ fm}^2,
\end{equation}

where $m_N$ is the nucleon mass and $N$, $Z$, and $A$, are the neutron number, proton number and mass number of ions, respectively. If the nucleus emits any number of neutrons (Xn), the total probability is given by
\begin{equation}
    P_{(Xn)}(b) = 1 - e^{-P^1_{(Xn)}(b)}.
\end{equation}

For cases where both nuclei emit neutrons simultaneously, the probability of mutual neutron emission (XnXn) is given by:
\begin{equation}
    P_{(XnXn)}(b) = [P_{(Xn)}(b)]^2.
\end{equation}

Similarly, the probability of mutual single-neutron emission (1n1n) is
\begin{equation}
    P_{(1n1n)}(b) = [P_{(1n)}(b)]^2.
\end{equation}

After obtaining probabilities of electron-positron pair production and the probability of mutual single-neutron emission (1n1n), we can write the total cross section for electron positron pair production with mutual nuclear excitation as
\begin{equation}
    \sigma_{e^+e^-}^{\text{GDR}} = 2\pi \int_{b_{\text{min}}}^{\infty} db \, b \, P_{e^+e^-}(b) [P_{(1n)}(b)]^2,
\end{equation}

where $b_{min}$ is the minimum impact parameter for the ultra-peripheral heavy ion collision and this is $b_{min} > R_{1}+R_{2}$ greater then sum of the radius of the colliding ions.

\section{Results}

We have calculated electron-positron pair productions accompanied with GDR at RHIC and LCH energies. We have used STAR restrictions for transverse momentum $p_{\perp} < 0.15 \,\,GeV/c$, rapidity $|y| < 1$ and invariant mass of pairs  $ M > 0.4 \,\, GeV/c^2$. We have then compare the calculations of RHIC and LHC energies. In Table \ref{table1}, we have tabulated unrestricted and restricted cross sections of colliding heavy ions. First column shows the cross sections of pure electron-positron pair productions. Second column shows the tagged 1n1n neutron emission and third column shows the XnXn neutron emissions. Calculations show that ratio of pure electron-positron pair production cross sections LHC (2.76 TeV) to RHIC (200 GeV) is almost 4. One the other hand, the same ratio for the neutron emissions is almost 2.

\begin{table}[H]
\tbl{Cross sections for Au + Au collisions at RHIC energies and for Pb + Pb collisions at LHC energies for electron-positron pair production accompanied by giant dipole resonance. \label{table1}}
{\begin{tabular}{@{}cccc@{}}  
        \toprule
         & $\sigma$  & $\sigma_{\text{tagged}}$ [1n1n]   & $\sigma_{\text{tagged}}$ [XnXn]  \\
         &(b)&(mb)&(mb) \\
        \midrule
        \multicolumn{4}{c}{\textbf{Au + Au at RHIC (200 GeV)}} \\
        Without restriction & 56029 & 421 & 544 \\
        With restriction & 0.26 & 1.50 & 2.04 \\
        \midrule
        \multicolumn{4}{c}{\textbf{Pb + Pb at LHC (2.76 TeV)}} \\
        Without restriction & 224008 & 897 & 1170 \\
        With restriction & 1.33 & 2.93 & 4.21 \\
        \bottomrule
    \end{tabular}}

\end{table}

In Fig.\ref{fig:rapgraph}, we have calculated the differential cross sections as a function of pair rapidity. Calculated differential cross sections are shown for the four collision systems for Au+Au at RHIC 200 GeV per nucleon with and without restrictions, and Pb+Pb at LHC 2.76 TeV per nucleon with and without restrictions. STAR restriction for the rapidity is $ |y| < 1$ and we have also restricted our calculation accordingly. The figures clearly shows that the differential cross sections as a function of pair rapidity increase with colliding energies of the nuclei.

\begin{figure}[H]
    \centering
    \begin{subfigure}[b]{0.45\textwidth}
        \centering
        \includegraphics[width=\textwidth]{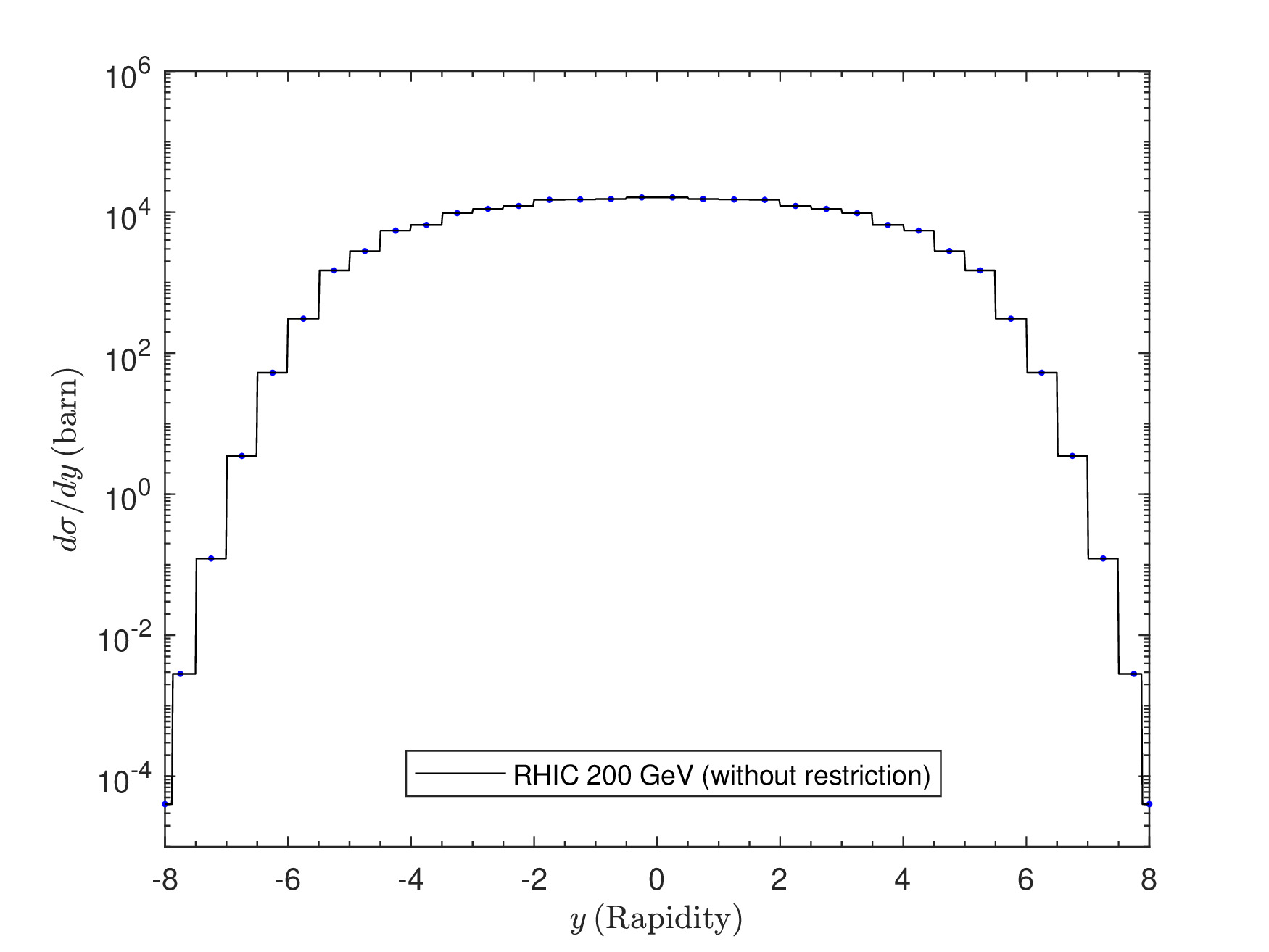}
        
        \caption{}
        \label{fig:grafik1}
    \end{subfigure}
    \hfill
    \begin{subfigure}[b]{0.45\textwidth}
        \centering
        \includegraphics[width=\textwidth]{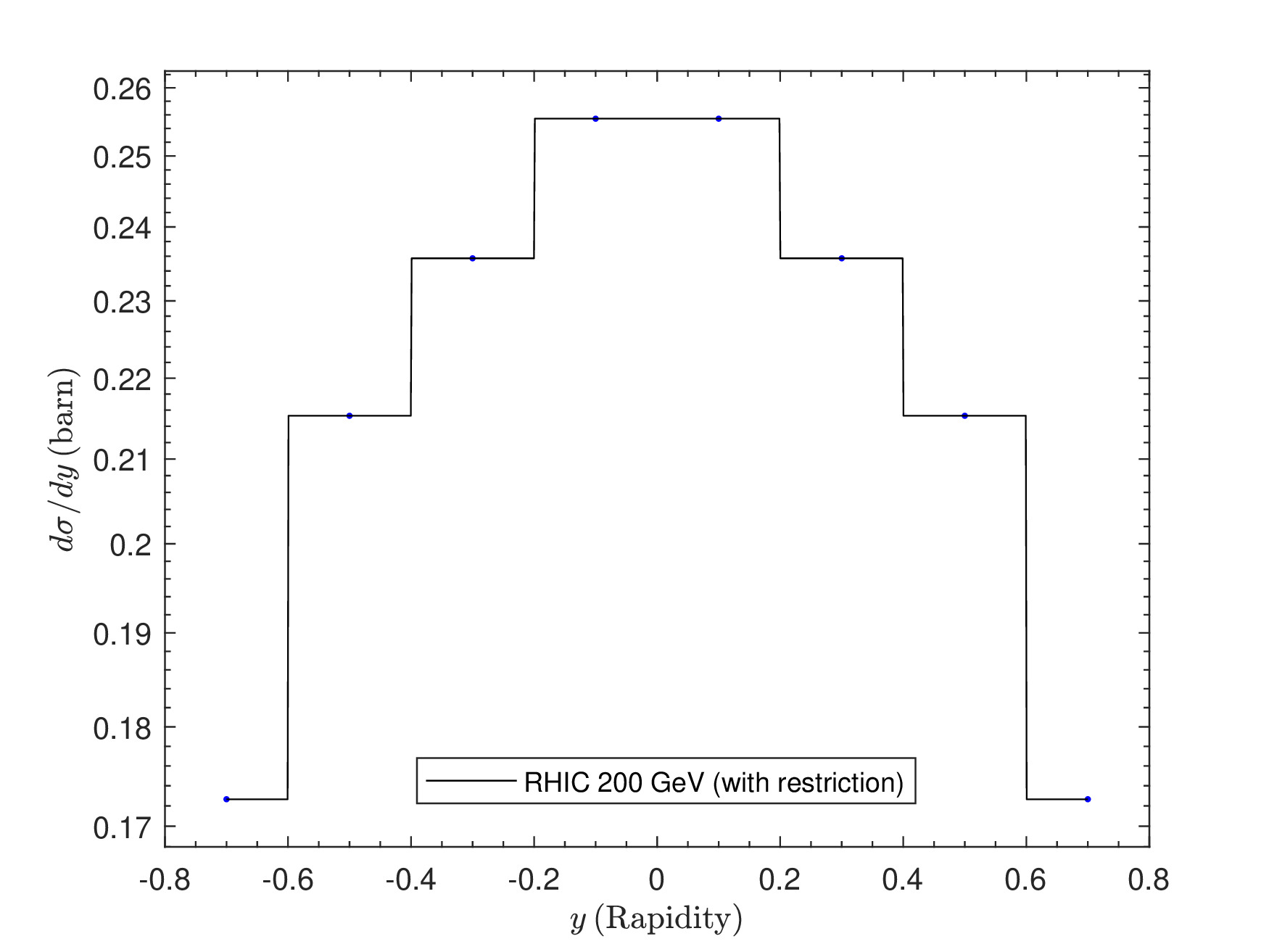}
        \caption{}
        \label{fig:grafik2}
    \end{subfigure}
    \vskip\baselineskip
    \begin{subfigure}[b]{0.45\textwidth}
        \centering
        \includegraphics[width=\textwidth]{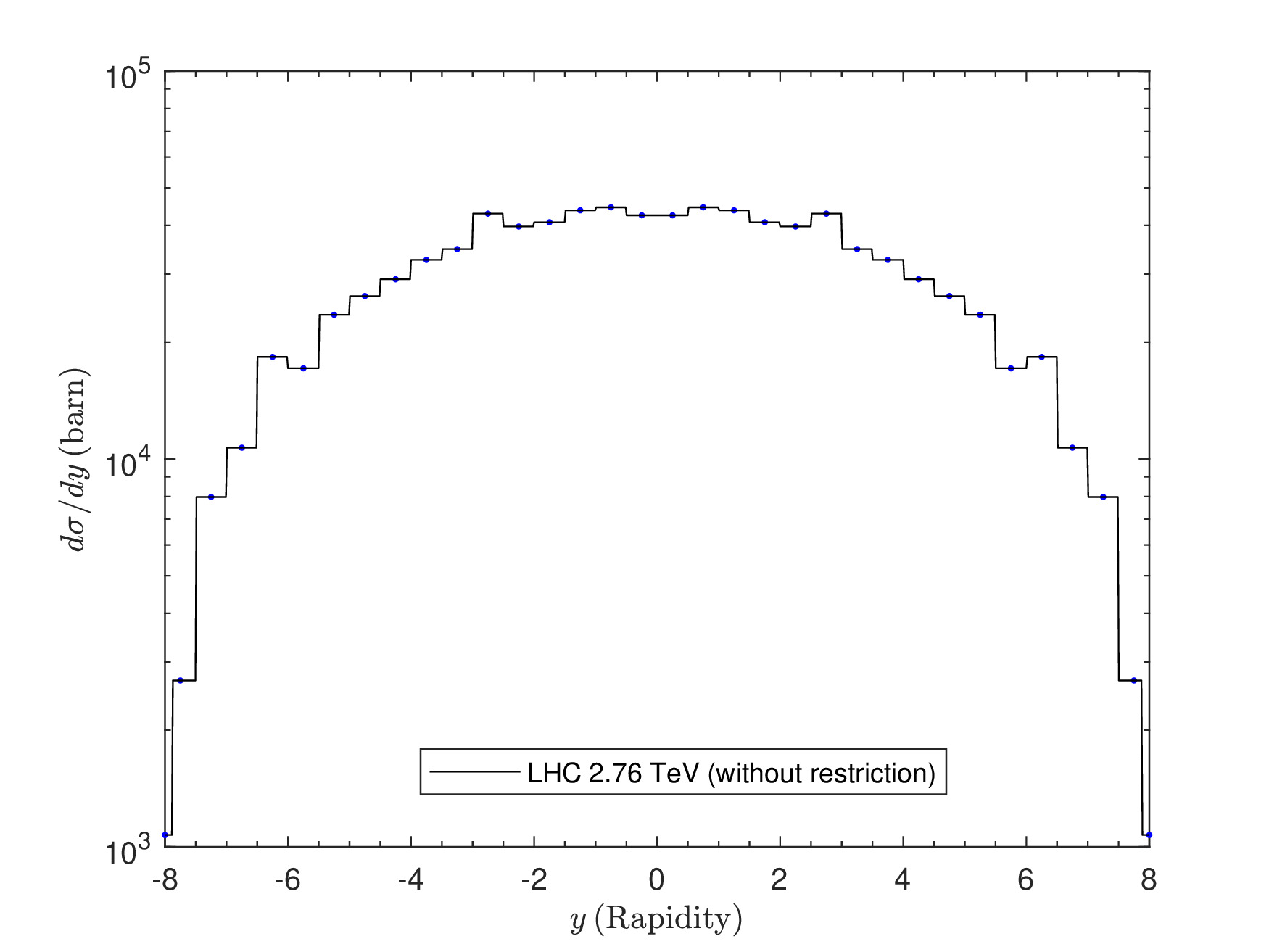}
        \caption{}
        \label{fig:grafik3}
    \end{subfigure}
    \hfill
    \begin{subfigure}[b]{0.45\textwidth}
        \centering
        \includegraphics[width=\textwidth]{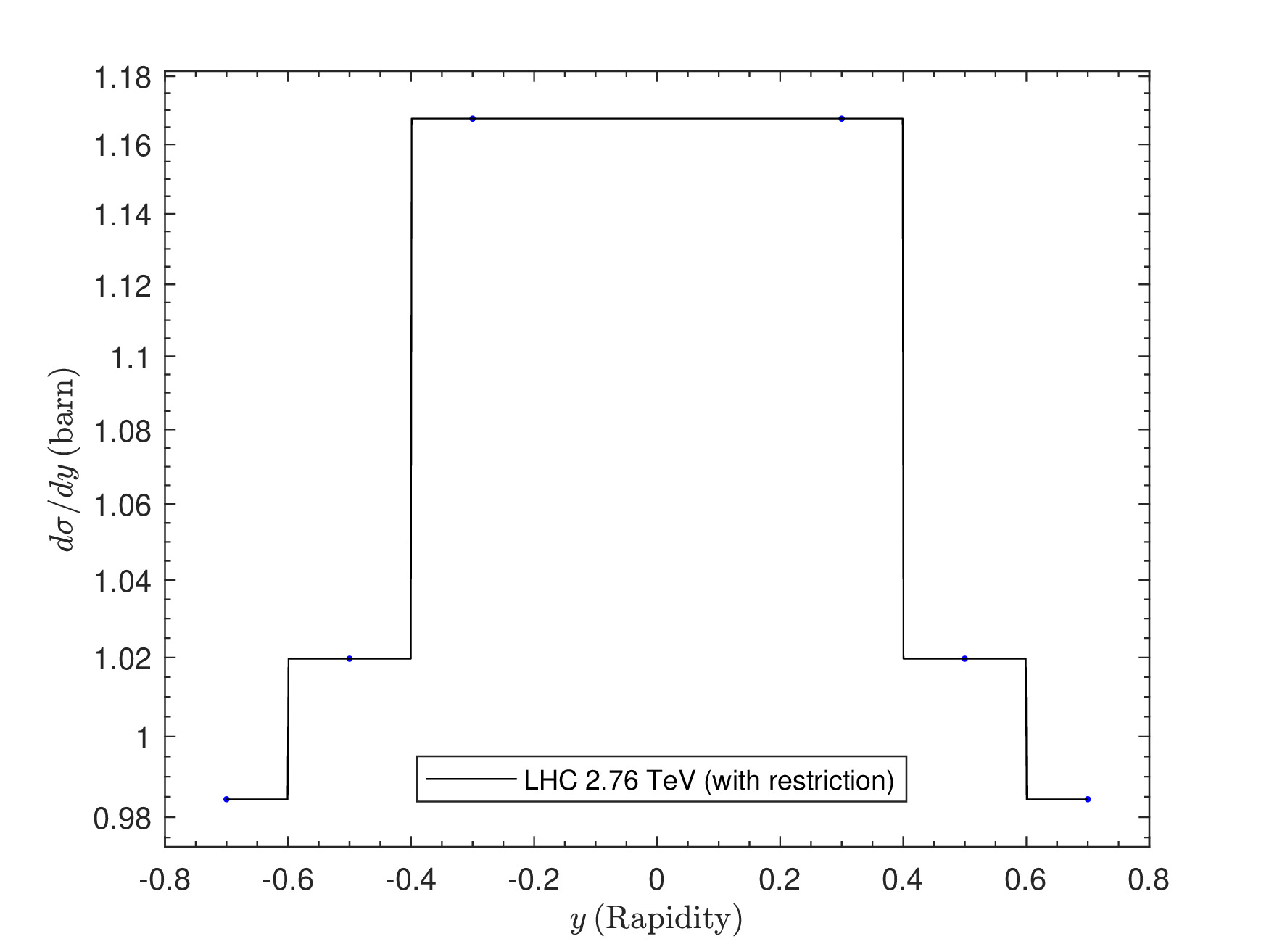}
        \caption{}
        \label{fig:rapgraph}
    \end{subfigure}
    \caption{Differential cross section as a function of the rapidiy ($y$) of the produced electron-positron pair. Calculated differential cross
sections are shown for the four collision systems (a) Au+Au at RHIC 200 GeV per nucleon without restriction, (b) Au+Au at RHIC 200 GeV per nucleon with restriction, (c) Pb+Pb at LHC 2.76 TeV per nucleon without restriction and (d) Pb+Pb at LHC 2.76 TeV per nucleon with restriction.}
    \label{fig:rapgraph}
\end{figure}

In Fig. \ref{fig:tmgraph}, we have plotted differential cross sections as a function of the transverse momentum ($p_\perp$) of the produced electron-positron pair. For the restricted case we have consider the pair transverse momentum $p_\perp < 0.15 \,\,GeV/c$.

\begin{figure}[H]
    \centering
    \begin{subfigure}[b]{0.45\textwidth}
        \centering
        \includegraphics[width=\textwidth]{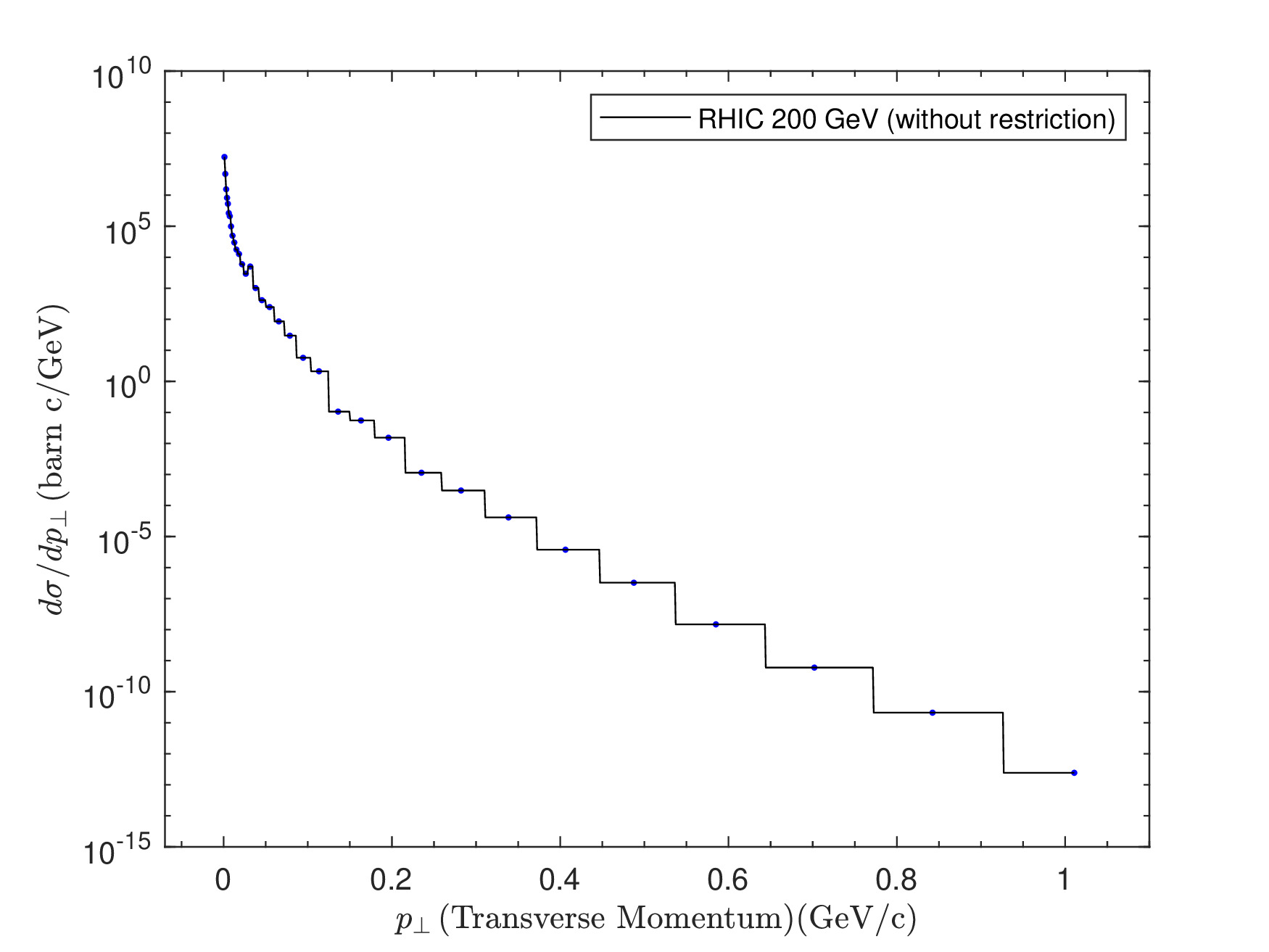}
        \caption{}
        \label{fig:grafik1}
    \end{subfigure}
    \hfill
    \begin{subfigure}[b]{0.45\textwidth}
        \centering
        \includegraphics[width=\textwidth]{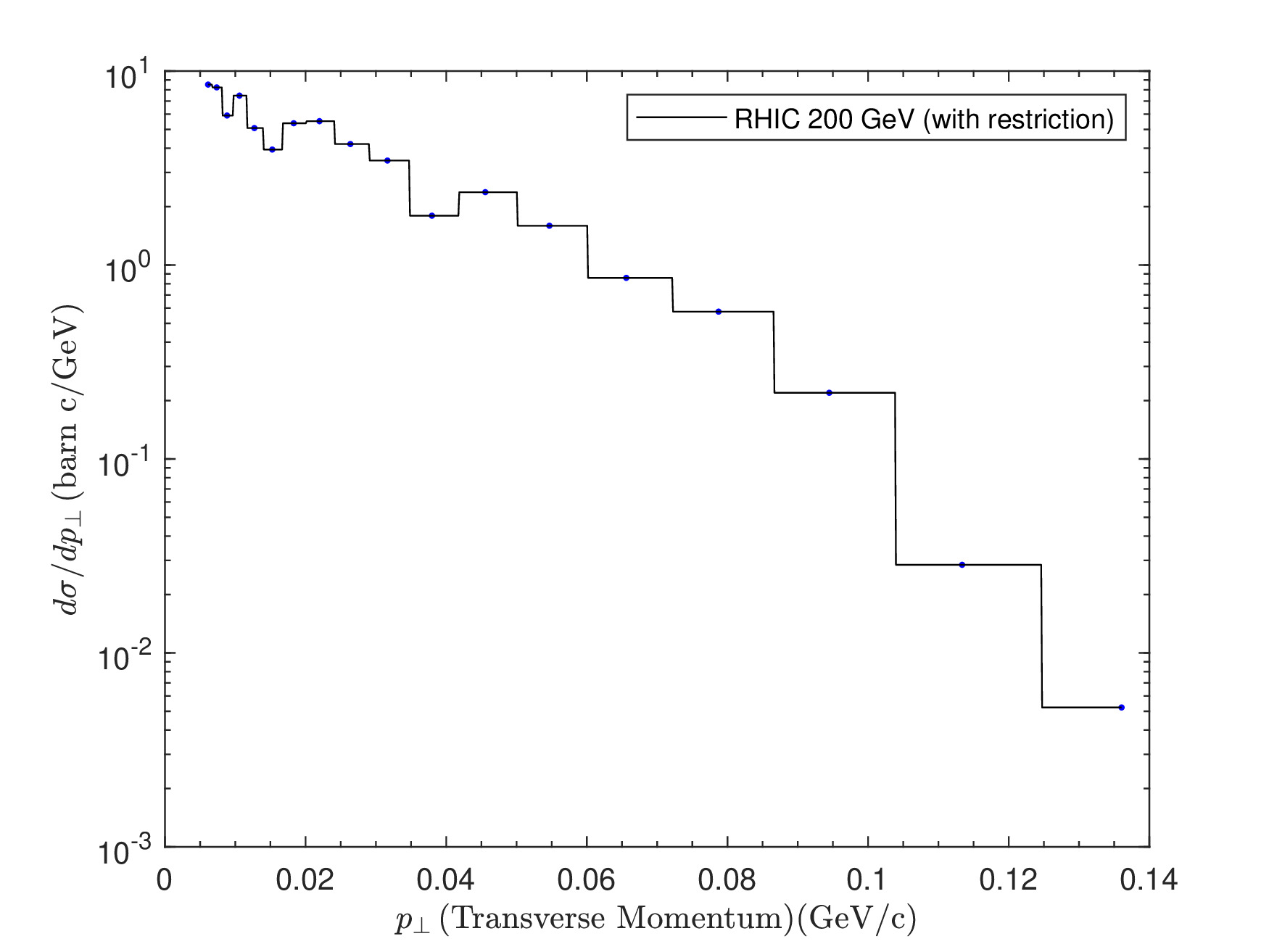}
        \caption{}
        \label{fig:grafik2}
    \end{subfigure}
    \vskip\baselineskip
    \begin{subfigure}[b]{0.45\textwidth}
        \centering
        \includegraphics[width=\textwidth]{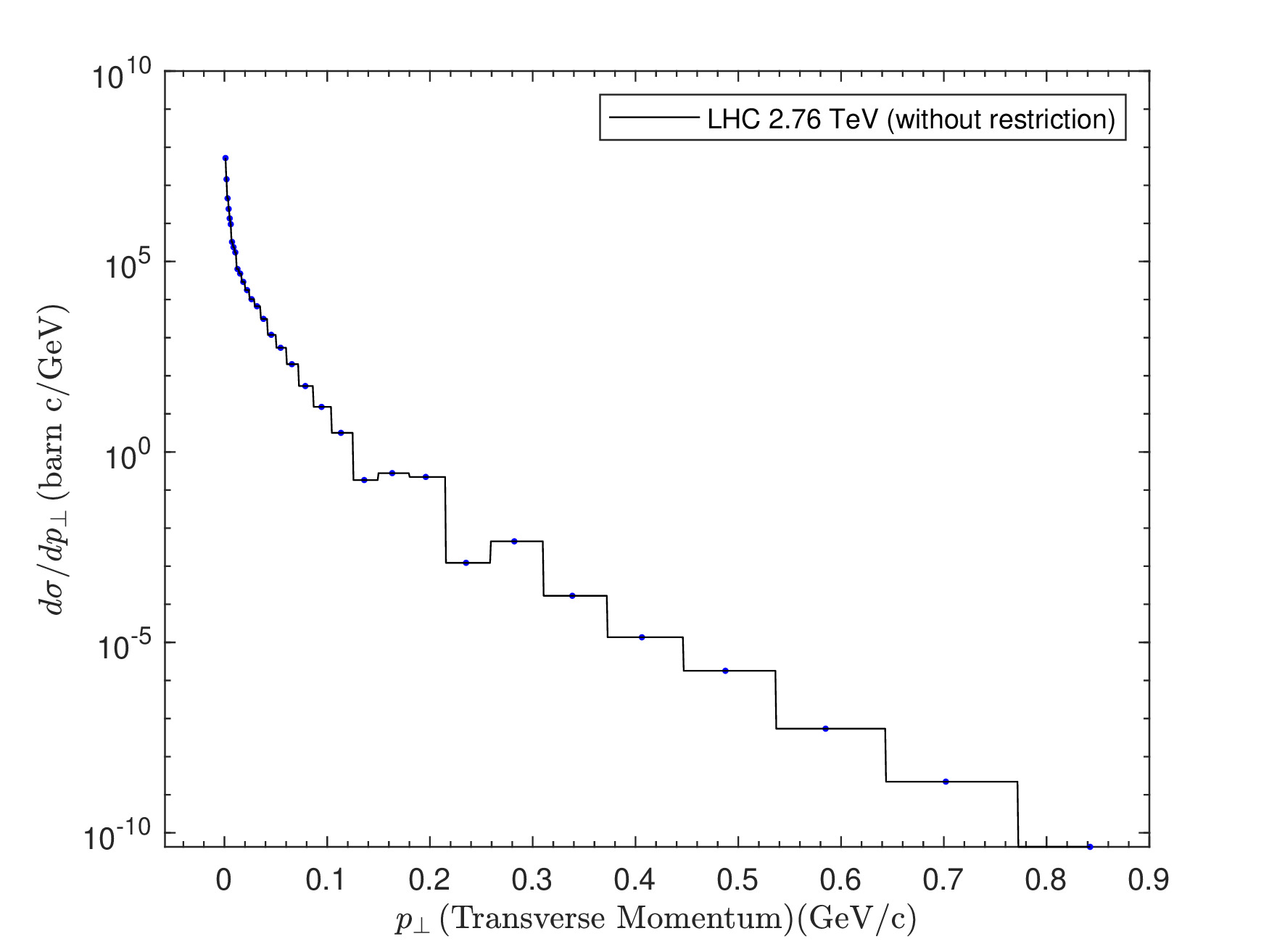}
        \caption{}
        \label{fig:grafik3}
    \end{subfigure}
    \hfill
    \begin{subfigure}[b]{0.45\textwidth}
        \centering
        \includegraphics[width=\textwidth]{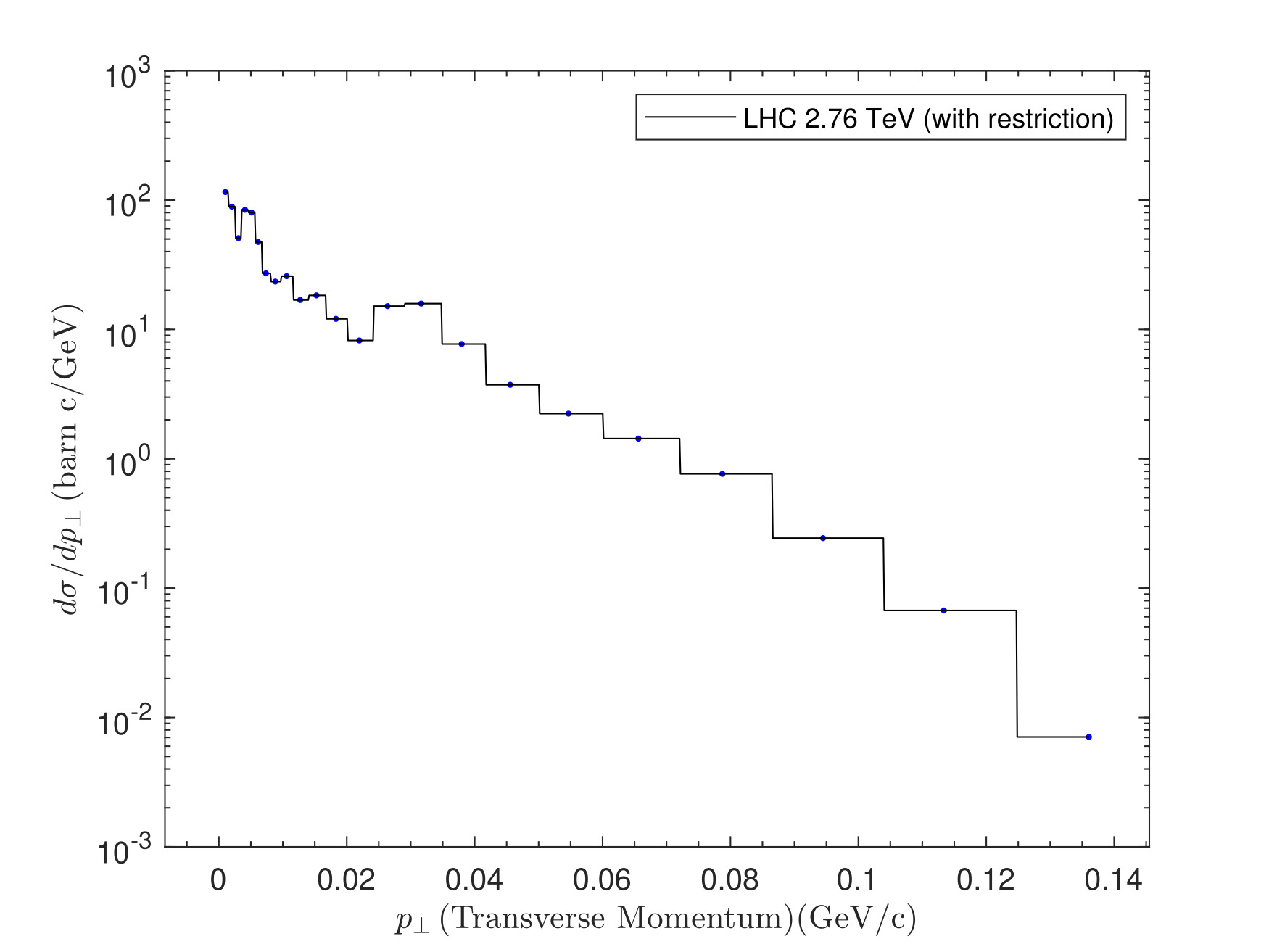}
        \caption{}
        \label{fig:tmgraph}
    \end{subfigure}
    \caption{Differential cross section as a function of the transverse momentum ($p_\perp$) of the produced electron-positron pair. Calculated differential cross
sections are shown for the four collision systems (a) Au+Au at RHIC 200 GeV per nucleon without restriction, (b) Au+Au at RHIC 200 GeV per nucleon with restriction, (c) Pb+Pb at LHC 2.76 TeV per nucleon without restriction and (d) Pb+Pb at LHC 2.76 TeV per nucleon with restriction.}
    \label{fig:tmgraph}
\end{figure}

In Fig. \ref{fig:massgraph}, we have plotted the differential cross sections as a function of the invariant mass ($M$) of the produced electron-positron pair. For the restricted case we have consider the pair invariant mass $M > 0.4 \,\,GeV/c^2$.

\begin{figure}[t]
    \centering
    \begin{subfigure}[b]{0.45\textwidth}
        \centering
        \includegraphics[width=\textwidth]{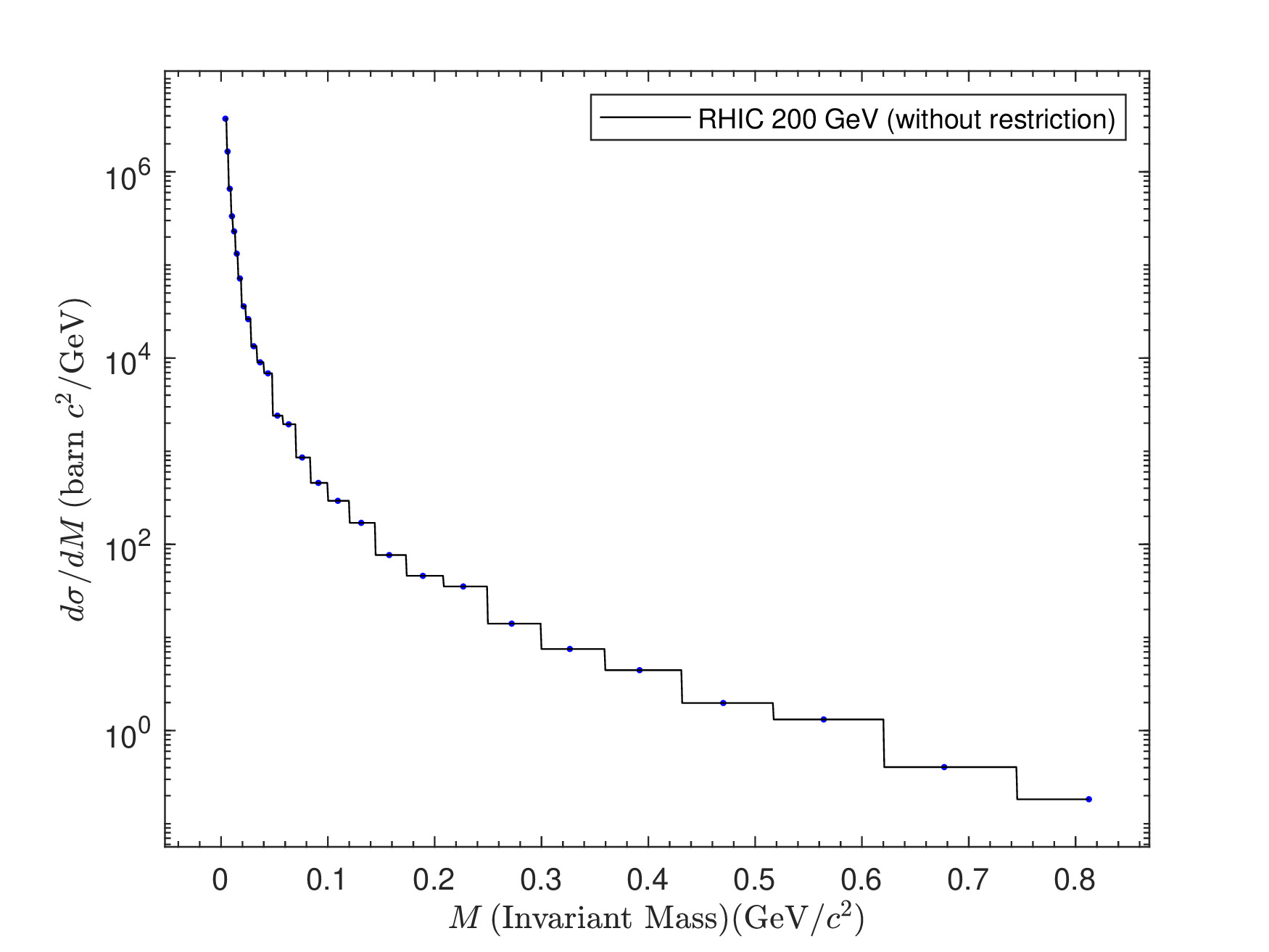}
        \caption{}
        \label{fig:grafik1}
    \end{subfigure}
    \hfill
    \begin{subfigure}[b]{0.45\textwidth}
        \centering
        \includegraphics[width=\textwidth]{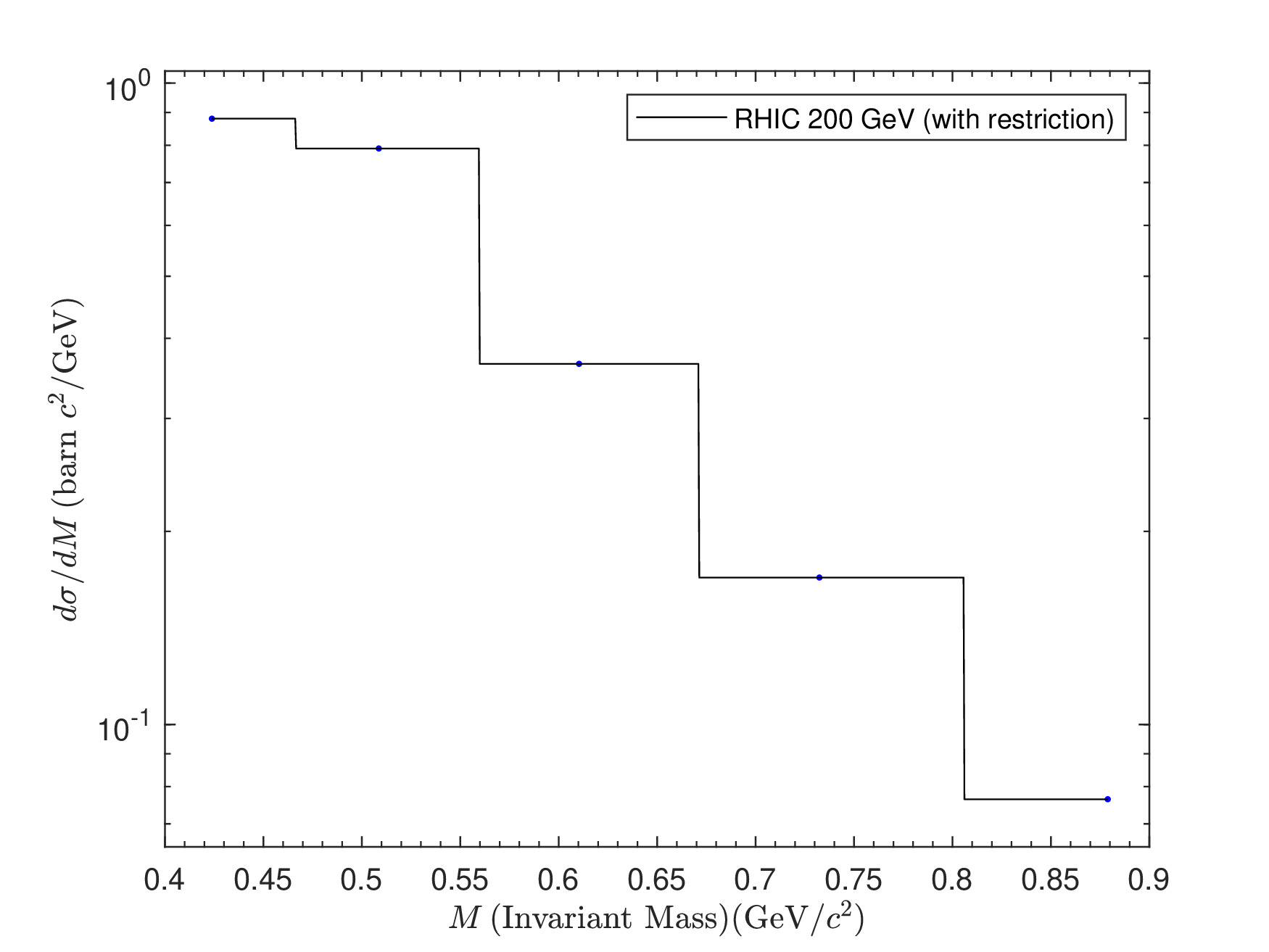}
        \caption{}
        \label{fig:grafik2}
    \end{subfigure}
    \vskip\baselineskip
    \begin{subfigure}[b]{0.45\textwidth}
        \centering
        \includegraphics[width=\textwidth]{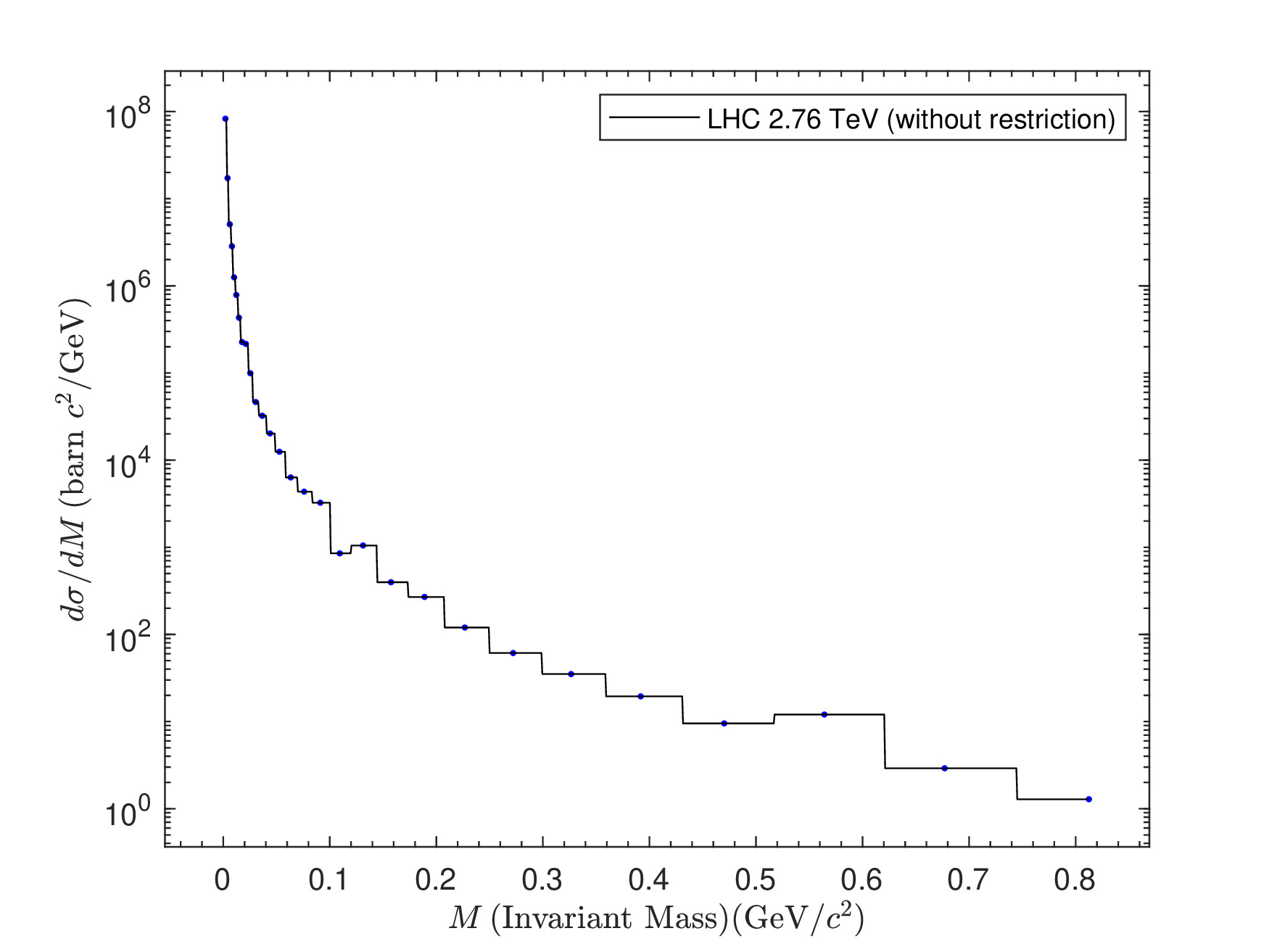}
        \caption{}
        \label{fig:grafik3}
    \end{subfigure}
    \hfill
    \begin{subfigure}[b]{0.45\textwidth}
        \centering
        \includegraphics[width=\textwidth]{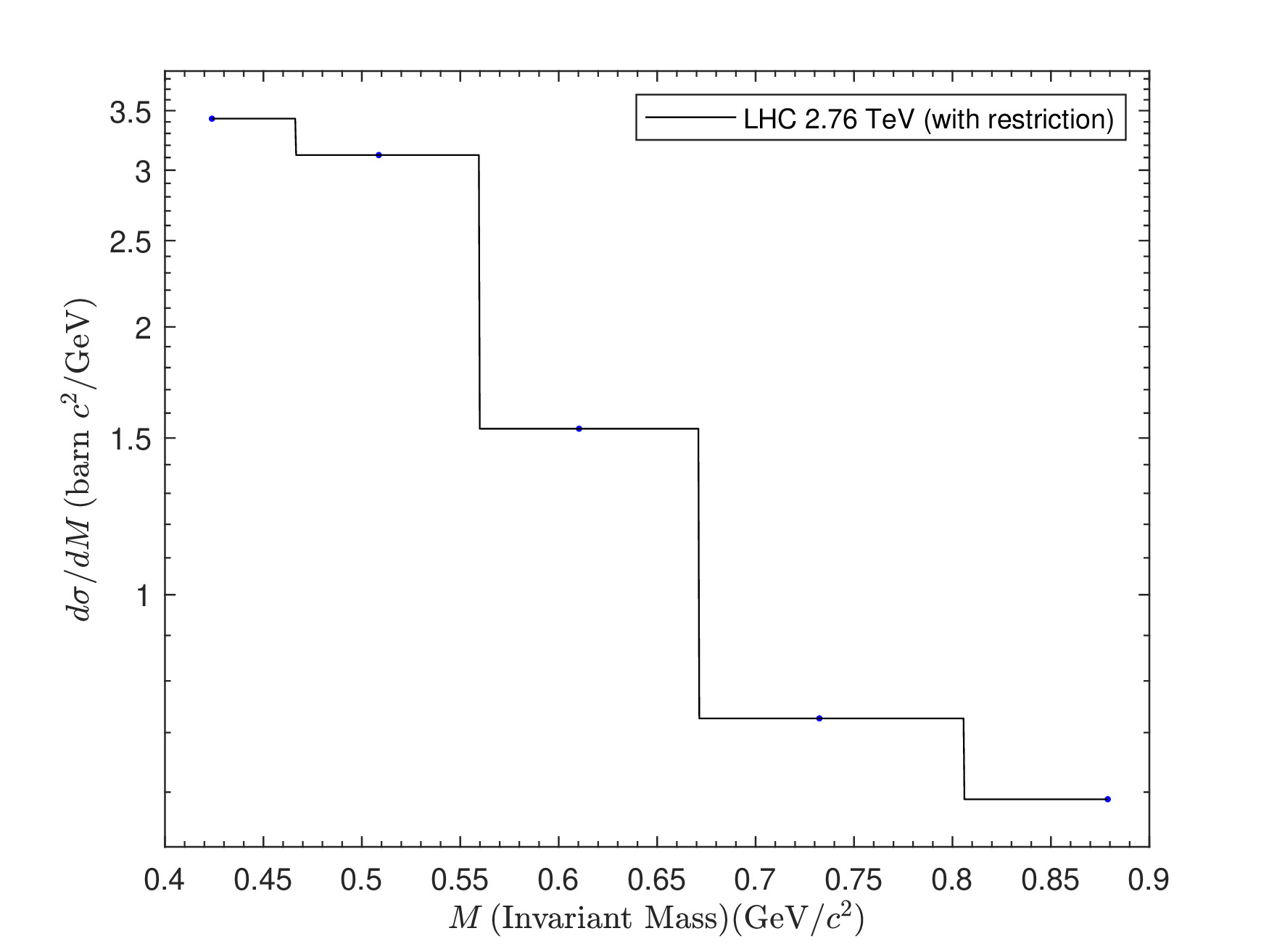}
        \caption{}
        \label{fig:massgraph}
    \end{subfigure}
    \caption{Differential cross section as a function of the invariant mass ($M$) of the produced electron-positron pair. Calculated differential cross sections are shown for the four collision systems (a) Au+Au at RHIC 200 GeV per nucleon without restriction, (b) Au+Au at RHIC 200 GeV per nucleon with restriction, (c) Pb+Pb at LHC 2.76 TeV per nucleon without restriction and (d) Pb+Pb at LHC 2.76 TeV per nucleon with restriction.}
    \label{fig:massgraph}
\end{figure}

\section{Conclusion}

We have calculated the electron-positron pair production cross section accompanied by GDR by using the Monte Carlo method. We have improved our previous calculations and find more accurate impact parameter dependence cross section. We have used STAR experimental restrictions in our calculations. In future experiments at RHIC and at LHC, there are some plans to improve the experimental conditions to measure the higher order effects in lepton pair productions. We can also investigate the higher order Coulomb effects in these collisions. The production of photo-nuclear interactions at LHC is much higher than the previous colliders. This will play important roles to the luminosity of the beam. Publications about peripheral relativistic heavy-ion collisions \cite{klein2019acoplanarity,klein2020lepton,klusek2021centrality,adam2021measurement} shows that the impact parameter dependence cross sections of lepton pair production are very important and detail knowledge of impact parameter dependence cross sections particularly for small impact parameters can help to understand many physical events in STAR and LHC experiments.

In this work, we have just calculated electron-positron pair production together with Coulomb excitation mainly GDR. However, colliding energies of heavy ions at LHC are very large so that the probabilities of producing heavy leptons such as muon and tauon pair are also large and they can be measured experimentally. In future work, we are planning to include the heavy leptons in our calculations together with GDR. These calculations and measurements provide us information to better understand the QED in a strong field.

\section*{ORCID}

\noindent Atacan Fatih Candar - \url{https://orcid.org/0009-0005-0074-7870}

\noindent Mehmet Cem Güçlü - \url{https://orcid.org/0000-0002-0764-084X}

\bibliographystyle{ws-ijmpa}
\bibliography{references}

\end{document}